\documentclass[matprg]{svjour}
\usepackage{graphics}
\usepackage{amssymb,amsmath,color,graphics,epsfig}
\newtheorem{fact}[theorem]{Fact}

\newcommand{\poly}{\textrm{poly}}

\newcommand{\conv}{\mathrm{conv}}

\newcommand{\OSOPTK}{\mathcal{O}_{\text{\begin{tiny}\textsc{SOPT}($K$)\end{tiny}}}}

\newcommand{\OSSEPKprime}{\mathcal{O}_{\text{\begin{tiny}\textsc{SSEP}($K'$)\end{tiny}}}}
\newcommand{\OSSEPKstar}{\mathcal{O}_{\text{\begin{tiny}\textsc{SSEP}($K^\star$)\end{tiny}}}}
\newcommand{\OSSEPQp}{\mathcal{O}_{\text{\begin{tiny}\textsc{SSEP}($Q_p$)\end{tiny}}}}

\begin{document}
\author{L. M. Ioannou\inst{1} \and B. C. Travaglione\inst{1}\inst{2} \and D. Cheung\inst{3}}
\title{Convex Separation from Optimization via Heuristics}
\institute{L. M. Ioannou \at Department of Applied Mathematics and
Theoretical Physics, Centre For Mathematical Sciences, University
of Cambridge, Wilberforce Road, Cambridge, CB3 0WA, U.K. \and B.
C. Travaglione \at Department of Applied Mathematics and
Theoretical Physics, Centre For Mathematical Sciences, University
of Cambridge, Wilberforce Road, Cambridge, CB3 0WA, U.K.; and
Computer Laboratory, University of Cambridge, J.J. Thomson Ave,
Cambridge CB3 0FD, U.K.\and D. Cheung \at Department of
Combinatorics and Optimization, Faculty of Mathematics, University
of Waterloo, Ontario, N2L 3G1, Canada.}
\date{Received: date / Revised version: date}
\maketitle
\begin{abstract}
Let $K$ be a full-dimensional convex subset of $\mathbb{R}^n$. We
describe a new polynomial-time Turing reduction from the weak
separation problem for $K$ to the weak optimization problem for
$K$ that is based on a geometric heuristic. We compare our
reduction, which relies on analytic centers, with the standard,
more general reduction.
\end{abstract}
\section{Introduction}\label{sec_Intro}

Let $K$ be a full-dimensional convex subset of $\mathbb{R}^n$
containing a ball of finite radius centered at $c_0$ and contained
in a ball of finite radius $R$. The \emph{separation problem} for
$K$ is that of finding a hyperplane that separates a given point
$p$ from $K$, or concluding that $p\in K$. The separation problem
is quite general in that it has been shown to be polynomial-time
equivalent to other natural convex programming problems
\cite{GLS88}.  One of these problems is the \emph{optimization
problem}, which is the problem of maximizing a linear functional
$c^Tx$ over all $x\in K$ for a given $c\in\mathbb{R}^n$.

It may arise in some applications that solving the optimization
problem for $K$ is more practically feasible than solving the
separation problem directly, for example, if the extreme points of
$K$ are parameterized by a number of parameters that is
significantly smaller than $n$.\footnote{Clearly this condition by
no means guarantees that the optimization problem is easier.
However, even if this is not the case, the optimization problem
may still be more ``feasible'' than the separation problem if only
because the former is better studied and thus software for it more
readily available.} In such a case, it might make sense to solve
an instance of the separation problem by solving polynomially many
instances of the optimization problem, using a Turing reduction
\cite{GJ79} from separation to optimization. Despite this
practical philosophy, which the authors have adopted with some
success to the (NP-hard \cite{Gur03}) quantum separability
problem\footnote{The quantum separability problem is related to
the problem of entanglement detection.  Entanglement is a
fundamental physical resource which plays a central role in
quantum computing and quantum cryptography \cite{NC00}.}
\cite{ITCE04,qphIoa05}, the main purpose of research into
polynomial-time reductions among convex body problems is
theoretical -- to uncover the intrinsic beauty of the subject of
convex programming algorithms, which historically marries physical
intuition and mathematical rigor in a unique way. Our
heuristics-based reduction is a fine example of this marriage.

\section{Convex Body Problems and Cutting-plane Feasibility Algorithms}\label{sec_DefinitionsAndBackground}

Let $S(K,\delta)$ denote the union of all balls of radius $\delta$
with centers belonging to $K$, and let $S(K,-\delta)$ denote the
union of all centers of all balls of radius $\delta$ contained in
$K$, where the balls are defined with respect to the Euclidean
norm. We now give the formal definitions of the basic convex body
problems that will concern us, taken from
\cite{GLS88}.\footnote{As a reminder, because computers use finite
representation of numbers, the problems are suitably weakened with
small accuracy parameters. As well, the rational field
$\mathbb{Q}$ is used instead of the real field $\mathbb{R}$, and
the $l_\infty$-norm (maximum norm) appears instead of the
$l_2$-norm (Euclidean norm); however, these technicalities will
not be carried through the paper.}

%

\begin{definition}[Weak Optimization Problem for $K$ (WOPT($K$))]\label{def_WOPT}
Given a rational vector $c\in\mathbb{R}^n$ and rational
$\epsilon>0$, either
\begin{itemize}
\item find a rational vector $y\in\mathbb{R}^n$ such that $y\in
S(K,\epsilon)$ and  $c^Tx\leq c^Ty +\epsilon$ for every $x\in K$;
or

\item assert that $S(K,-\epsilon)$ is
empty.\label{eqn_WSEPEntAssertion}
\end{itemize}
\end{definition}

\begin{definition}[Weak Separation Problem for $K$ ($\textrm{WSEP($K$)}$)]\label{def_WSEP}
Given a rational vector $p\in\mathbb{R}^n$ and rational
$\delta>0$, either
\begin{itemize}
\item assert $p\in S(K,\delta)$,
\hspace{2mm}\textrm{or}\label{eqn_WSEPSepAssertion}

\item find a rational vector $c\in\mathbb{R}^n$ with
$||c||_\infty= 1$ such that $c^Tx < c^Tp$ for every $x\in
K$.\label{eqn_WSEPEntAssertion}
\end{itemize}
\end{definition}

\noindent Let $K'$ be a full-dimensional bounded convex subset of
$\mathbb{R}^n$.

\begin{definition}[Weak Violation Problem for $K'$ (WVIOL($K'$))]\label{def_WVIOL}
Given a rational vector $c\in\mathbb{R}^n$ and rationals $\gamma$
and $\epsilon'>0$, either
\begin{itemize}
\item assert that $c^Tx\leq\gamma+\epsilon'$ for all $x\in
S(K',-\epsilon')$, or

\item find a vector $y\in S(K',\epsilon')$ with
$c^Ty\geq\gamma-\epsilon'$.
\end{itemize}
\end{definition}

%
%

\noindent If $\gamma=-1$ and $c$ is the origin
$\bar{0}\in\mathbb{R}^n$, then WVIOL($K'$) reduces to the
\emph{weak feasibility problem for} $K'$, denoted WFEAS($K'$).  By
taking $\delta$, $\epsilon$, and $\epsilon'$ to be zero, we
implicitly define the corresponding \emph{strong} problems,
denoted SSEP, SOPT, SVIOL, and SFEAS.  Note that SFEAS($K'$) is
the problem of deciding whether $K'$ is empty or finding a point
in $K'$.

Central to our reduction are cutting-plane algorithms for
WFEAS($K'$), relative to a WSEP($K'$) oracle $\OSSEPKprime$.  All
such algorithms have the same basic structure:

\begin{enumerate}
\item Define a (possibly very large) regular bounded convex set
$P_0$ which is guaranteed to contain $K'$, such that, for some
reasonable definition of ``center'', the center $\omega_0$ of
$P_0$ is easily computed.  The set $P_0$ is called an \emph{outer
approximation} to $K'$.  Common choices for $P_0$ are the
origin-centered hyperbox, $\{x\in\mathbb{R}^n: -2^L \leq x_i \leq
2^L, \hspace{2mm}1\leq i\leq n \}$ and the origin-centered
hyperball, $\{x: x^Tx \leq 2^L\}$ (where $2^L$ is a trivially
large bound).

\item Give the center $\omega$ of the current outer approximation
$P$ to $\OSSEPKprime$.

\item If $\OSSEPKprime$ asserts ``$\omega\in K'$'', then HALT.

\item Otherwise, say $\OSSEPKprime$ returns the hyperplane
$\pi_{c,b}$ such that $K'\subset \{x: c^Tx \leq b\}$.  Update
(shrink) the outer approximation $P:=P\cap \{x: c^Tx \leq b'\}$
for some $b'\geq b$. Possibly perform other computations to
further update $P$. Check stopping conditions; if they are met,
then HALT. Otherwise, go to step (ii).
\end{enumerate}
\noindent  The difficulty with such algorithms is knowing when to
halt in step (iv).  Generally, the stopping conditions are related
to the size of the current outer approximation.  Because it is
always an approximate (weak) feasibility problem that is solved,
the associated accuracy parameter $\epsilon'$ can be exploited to
get a ``lower bound'' $V$ on the ``size'' of $K'$, with the
understanding that if $K'$ is smaller than this bound, then the
algorithm can correctly assert that $S(K',-\epsilon')$ is empty.
Thus the algorithm stops in step (iv) when the current outer
approximation is smaller than $V$.

The cutting-plane algorithm is called \emph{(oracle-)
polynomial-time} if it runs in time
$O(\poly(n,\log(R'/\epsilon')))$ with unit cost for the oracle,
where $R'$ is the outer radius of $K'$. It is called
\emph{(oracle-) fully polynomial} if it runs in time
$O(\poly(n,R'/\epsilon'))$.

There are a number of such polynomial-time convex feasibility
algorithms (see \cite{AV95} for a discussion of all of them).  The
three most important are the \emph{ellipsoid method}, the
\emph{volumetric center method}, and the \emph{analytic center
method}.  The ellipsoid method has $P_0=\{x:x^Tx\leq 2^L\}$ and is
the only one which requires ``further update'' of the outer
approximation $P$ in step (iv) after a cut has been made -- a new
minimal-volume ellipse is drawn around $P:=P\cap \{x: c^Tx \leq
b'\}$.  The ellipsoid method, unfortunately, suffers badly from
gigantic precision requirements, making it practically unusable.
The volumetric center and analytic center algorithms are more
efficient than the ellipsoid algorithm and are very similar to
each other in complexity and precision requirements, with the
analytic center algorithm having some supposed practical
advantages.\footnote{To date, no one has implemented a
polynomial-time cutting plane algorithm. For an implementation of
a fully polynomial algorithm, see
http://ecolu-info.unige.ch/logilab.}

The cutting plane $\{x: c^Tx = b'\}$ requires further definition:
\begin{equation}
\textrm{If } \left\{ \begin{array}{rcl}  b'&<& c^T\omega\\
b'&=& c^T\omega
\\  b'&>& c^T\omega\end{array}\right\} \textrm{ then the above is a } \left\{ \begin{array}{c} \textrm{\emph{deep-cut}} \\
\textrm{\emph{central-cut}}
\\ \textrm{\emph{shallow-cut}} \end{array}\right\}\textrm{ algorithm.}
\end{equation}
\noindent Intuitively, deep-cut algorithms should be fastest.
Ironically, though, except for the case of ellipsoidal algorithms
(which are practically inefficient), the algorithms that are
provably polynomial-time are central- or even shallow-cut
algorithms.

\section{Reduction via Heuristics}\label{sec_SearchHeuristic}

Assume without loss that $c_0=\bar{0}$. Our goal is to develop a
new oracle-polynomial-time algorithm for WSEP($K$) with respect to
an oracle for WOPT($K$) (running in time
$O(\poly(n,\log(R/\delta)))$).  We do this via a geometric
approach, which differs from the standard approach covered in the
next section.  In this section, we ignore the weakness of the
separation and optimization problems, as it obfuscates the main
ideas; that is, we assume we are solving SSEP($K$) with an oracle
for SOPT($K$).

\begin{figure}[ht]
\centering \resizebox{150mm}{!}{\includegraphics{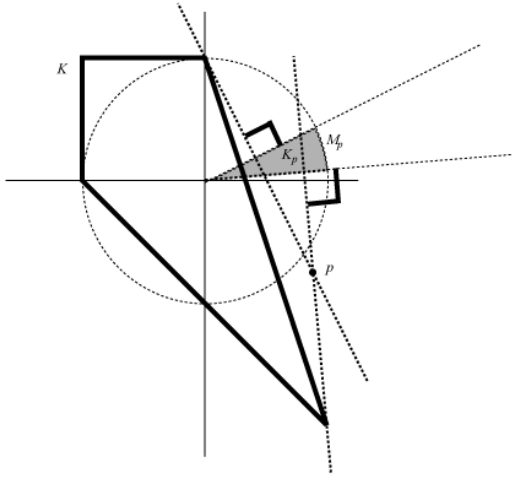}}
\caption[The sets $M_p$ and $K_p$ in $\mathbb{R}^2$]{The sets
$M_p$ and $K_p$ in $\mathbb{R}^2$.  Pictured in heavy outline is a
set $K$ in $\mathbb{R}^2$, where $K:=\conv\{(0,1), (-1,1), (-1,0),
(1,-2)\}$.  A point $p=(-7/8,-3/4)$ is shown as a heavy dot.  The
unit circle is drawn in a dashed line.  The set $M_p$ is the arc
of the unit circle that the shaded pie-slice subtends; the set
$K_p$ is the shaded pie-slice.  In two dimensions, the set $M_p$
($K_p$) is easy to construct.  This construction has been
illustrated: draw the two distinct lines through $p$ that are
tangent to $K$; the lines that determine the pie-slice are the two
straight lines that are perpendicular to the lines through $p$.}
\label{K_p}
\end{figure}

Suppose we have an oracle $\OSOPTK$ for the optimization problem
over $K$ such that, given a nonzero input vector $c$, $\OSOPTK$
outputs a point  $\OSOPTK(c)\\ \equiv k_c\in K$ that maximizes
$c^Tx$ for all $x\in K$. An important step in developing the
algorithm is noting that, given $\OSOPTK$, the search for a
separating hyperplane reduces to the search for a region on the
$(n-1)$-dimensional surface of the unit hypersphere $S_n$
(embedded in $\mathbb{R}^n$) centered at the origin. For $p\notin
K$, this region $M_p$ is simply $\{c\in S_n:\hspace{2mm} c^T
k_c<c^T p\}$ (see Figure \ref{K_p}).

The first observation is that, since $K$ properly contains the
origin, $M_p$ is contained in the hemisphere defined by $\{x: p^T
x\geq 0\}$:
\begin{fact}\label{Fact_InitialCut} For all $m\in M_p$, $m^Tp>0$.
\end{fact}
\begin{proof}
Let $m\in M_p$.  Then $m^Tp>m^Tk$ for all $k\in K$.  But the fact
that the 0-vector is properly contained in $K$ implies that there
exists $k\in K$ such that $m^Tk>0$.\end{proof}

\begin{figure}[ht]
\centering \resizebox{150mm}{!}{\includegraphics{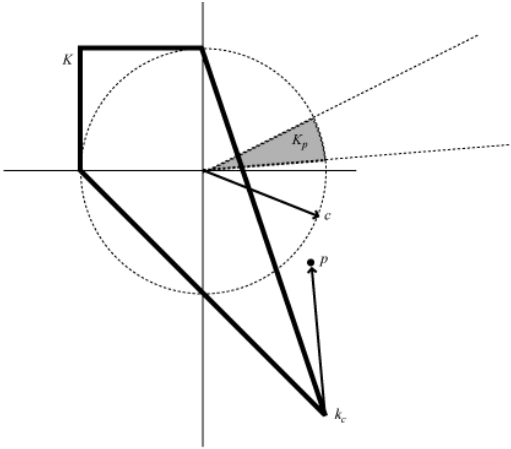}}
\caption[Illustration of heuristic behind Lemma
\ref{lem_Donny}]{Illustration of heuristic behind Lemma
\ref{lem_Donny}. Continuing from Figure \ref{K_p}, the unit vector
$c$ is a test vector that is close to $M_p$ but not in $M_p$.
Evidently, adding a component of $(p-k_c)$ to $c$ moves it closer
to $M_p$.  This intuition also holds in $\mathbb{R}^3$.}
\label{DonnyLemma}
\end{figure}

The second observation, Lemma \ref{lem_Donny}, is based on the
following heuristic, which can be pictured in $\mathbb{R}^2$ and
$\mathbb{R}^3$. Suppose $c$, $||c||=1$, is not in $M_p$ (but is
reasonably close to $M_p$) and that the oracle returns $k_c$. What
is a natural way to modify the vector $c$, so that it gets closer
to $M_p$? Intuition dictates moving $c$ \emph{away from} $k_c$ and
\emph{towards} $p$, that is, add a small component of the vector
$(p-k_c)$ to $c$, in order to generate a new guess
$c'=c+\lambda(p-k_c)/||p-k_c||$, for some $\lambda>0$, which we
could then give to the oracle again (see Figure \ref{DonnyLemma}).
Incidentally, we have found that the following little program,
which embodies this heuristic, actually performs well:\footnote{We
tested the program in $\mathbb{R}^{15}$, where $K$ is the convex
hull of $\{\alpha\alpha^*\otimes\beta\beta^*:
\alpha,\beta\in\mathbb{C}^2, ||\alpha||_2=||\beta||_2=1\}$ (which
is viewed as full-dimensional in $\mathbb{R}^{15}$), where $*$
denotes complex-conjugate transpose, and $\otimes$ denotes
Kronecker product.  This corresponds to the quantum separability
problem for two qubits \cite{qphIoa05}.}

\vspace{2mm} \fbox{\begin{minipage}{11cm}
\begin{description} \item
$c:=p/||p||;\hspace{2mm}d:=1;\hspace{2mm}i:=0;$ \item
\textsc{while}\hspace{2mm}$(d>0$\hspace{2mm}\textsc{and}\hspace{2mm}$i<N)$\hspace{2mm}\textsc{do}\hspace{2mm}$\lbrace$
\item\hspace{5mm}$k_c:=\OSOPTK(c);$
\item\hspace{5mm}$d:=c^Tk_c-c^Tp;$
\item\hspace{5mm}\textsc{if}$\hspace{2mm}(d<0)$\hspace{2mm}\textsc{then}\hspace{2mm}$\lbrace$
\textsc{return} $c$ $\rbrace$
\item\hspace{10mm}\textrm{\textsc{else}\hspace{2mm}$\lbrace$ $c:=
c+d(p-k_c)/||p-k_c||$};\hspace{2mm}$c:=c/||c||$;\hspace{2mm}$i:=i+1\rbrace\rbrace;$
\item\textrm{\textsc{return} ``INCONCLUSIVE''}
\end{description}
\end{minipage}}
\vspace{2mm}

\noindent This program can be regarded as an extremely simple
heuristic for the separation problem when given an optimization
subroutine (of course, it may give inconclusive results; in
practice, one should set $N$ as large as is practically feasible).

Interestingly, the above heuristic can be formalized as follows.
If $c$ is not in $M_p$ but is sufficiently close to $M_p$, then
$c$, $p$, and $k_c$ can be used to define a hemisphere which
contains $M_p$ and whose great circle cuts through $c$. More
precisely:
\begin{lemma}\label{lem_Donny} Suppose $m\in M_p$,
$c\notin M_p$, and let $\bar{a}:=(p-k_c)-\mathrm{Proj}_c(p-k_c)$.
If $m^Tc\geq 0$ then $m^T\bar{a}> 0$.
\end{lemma}
\begin{proof}
Note that $m^T\bar{a} = m^T(p-k_c)-[c^T(p-k_c)](m^Tc)$.  The
hypotheses of the lemma immediately imply that $m^T(p-k_c)>0$ and
$c^T(p-k_c)\leq 0$.  Thus, if $m^Tc\geq 0$, then $m^T\bar{a}> 0$.
\end{proof}
\noindent The lemma gives a method for reducing the search space
after each query to $\OSOPTK$ by giving a cutting plane
$\{x:\bar{a}^Tx=0\}$ that slices off a portion of the search
space.
Our search problem can easily be reduced to an instance of
SFEAS($K'$), where
$K'$ is 
\begin{eqnarray}\label{def_K_p}
K_p:=\left[\mathrm{ConvexHull}\left( M_p\cup \{\bar{0}\}\right)
\right] \setminus \{\bar{0}\}.
\end{eqnarray}
The set $K_p$, if not empty, can be viewed as a cone-like object,
emanating from the origin and cut off by the unit hypersphere (see
Figure \ref{K_p}).
The oracle $\OSOPTK$, along with Lemma \ref{lem_Donny},
essentially gives a separation oracle for $K_p$, as long as the
test vectors $c$ given to $\OSOPTK$ satisfy
\begin{eqnarray}\label{HypothesisOfDonnyLemma}
\text{$m^T c\geq 0$ for all $m\in M_p$}.
\end{eqnarray}


How could we ensure that all our test vectors $c$ satisfy
(\ref{HypothesisOfDonnyLemma})?  Recall Fact
\ref{Fact_InitialCut}, which says that the set $K_p$ is contained
in the halfspace $\lbrace x:p^T x\geq 0\rbrace$. Let $B_n$ be the
origin-centered unit hyperball in $\mathbb{R}^n$ and let
$a_1:=p/||p||$. Thus, straight away, the search space is reduced
to the hemisphere $B_n\cap\lbrace x:a_1^T x\geq 0\rbrace$. The
first test vector to give to the oracle $\OSOPTK$ is $p/||p||$,
which clearly has nonnegative dot-product with all points in $K_p$
and hence all $m\in M_p$.  By way of induction, assume that, at
some later stage in the algorithm, the current search space has
been reduced to $P:=B_n\bigcap \cap_{i=1}^{h}\lbrace x:a_i^T x\geq
b_i \rbrace$ by the generation of cutting planes $\lbrace x: a_i^T
x=b_i \rbrace$, where the $a_i$, for $i=2,3,\ldots, h$, are the
normalized $\bar{a}$ from $h-1$ invocations of Lemma
\ref{lem_Donny}.  Let $\omega$ be the ``center'' of $P$, and
\emph{suppose} that this ``center'' is a positive linear (conic)
combination of the normal vectors $a_i$, that is,
\begin{eqnarray}\label{eqn_CenterIsConicCombinationOfNormals}
\omega = \sum_{i=1}^h \lambda_i a_i,\hspace{2mm}\textrm{where
$\lambda_i\geq 0$ for all $i=1,2,\ldots, h$.}
\end{eqnarray}
Then, by inductive hypothesis, this implies that $m^T \omega\geq
0$ for all $m\in M_p$. Thus, $c:=\omega/||\omega||$ is a suitable
vector to give to the oracle $\OSOPTK$ and use in Lemma
\ref{lem_Donny}. Therefore, it suffices to find a definition of
``center $\omega$ of $P$'' that satisfies
(\ref{eqn_CenterIsConicCombinationOfNormals}), in order that all
our test vectors $c$ satisfy (\ref{HypothesisOfDonnyLemma}).
 Because we require (\ref{HypothesisOfDonnyLemma}), none of the
feasibility algorithms mentioned in Section
\ref{sec_DefinitionsAndBackground} can be applied directly.
However, the analytic-center algorithm due to Atkinson and Vaidya
\cite{AV95} beautifully lends itself to
a modification that allows (\ref{HypothesisOfDonnyLemma}) to be satisfied.  

Reducing the separation problem for $K$ to the convex feasibility
problem for some $K'$, while using the optimization oracle for $K$
as a separation oracle for $K'$, is not a new concept in convex
analysis.  But the precise way that Lemma \ref{lem_Donny}
generates each new cutting plane, incorporating the
correction-heuristic, does not appear in the literature.  This is
likely because there is a more general way to carry out such a
reduction, covered in the next section, which does not require
(\ref{eqn_CenterIsConicCombinationOfNormals}).\footnote{The more
general reduction was established by the early 1980s. Our
reduction seems to require the result in \cite{AV95}, which was
not known prior to 1992.}

\section{Comparison to Standard Reduction}\label{sec_ConnToOtherKnownMethods}

Note that our reduction requires knowledge of $c_0$ and $R$. A
polynomial-time Turing reduction from WSEP($K$) to WOPT($K$) is
well known -- even when $c_0$ and $R$ are unknown. One such
(ellipsoidal) reduction is given in Theorem 4.4.7 in \cite{GLS88};
however, it may be more general than is necessary. If $R$ is
known, then the standard way to perform the reduction of the
previous section may be found in the synthesis of Lemma 4.4.2 and
Theorem 4.2.2 in \cite{GLS88}; we outline it in the following
paragraph.

\begin{definition}[Polar of $K$] The \emph{polar} $K^\star$ of a full-dimensional convex set
$K\subset \mathbb{R}^n$ that contains the origin is defined as
\begin{eqnarray}
K^\star := \{c\in\mathbb{R}^n: c^Tx\leq 1 \hspace{2mm}\forall x
\in K\}.
\end{eqnarray}
\end{definition}

If $c\in K^\star$, then the plane $\pi_{c,1}\equiv\{x: c^Tx=1\}$
separates $p\in\mathbb{R}^n$ from $K$ when $c^Tp >1$. Thus, the
separation problem for $p$ is equivalent to the feasibility
problem for $Q_p$, defined as\footnote{Note that $Q_p$ is
guaranteed not to be empty when $p\notin K$.  For, then, there
certainly exists some plane $\pi_{c',b'}$ separating $p$ from $K$.
But since $K$ contains the origin, $b'$ may be taken to be
positive. Thus $\pi_{c'/b', 1}$ separates $p$ from $K$.}
\begin{eqnarray}
Q_p := K^\star \cap \{c: p^Tc \geq 1\}.
\end{eqnarray}
As mentioned in the previous section (and elaborated on in the
next section), to solve the feasibility problem for any $K'$, it
suffices to have a separation routine for $K'$.  Because we can
easily build a separation routine $\OSSEPQp$ for $Q_p$ out of
$\OSSEPKstar$, it suffices to have a separation routine
$\OSSEPKstar$ for $K^\star$ in order to solve the feasibility
problem for $Q_p$.\footnote{We slightly abuse the oracular
``$\mathcal{O}$'' notation by using it for both truly oracular
(black-boxed) routines and for other (possibly not completely
black-boxed) routines.}  Building $\OSSEPQp$ out of
$\OSSEPKstar$ is done as follows:\\

\indent\indent\indent\indent\indent\indent\indent Routine $\OSSEPQp(y)$:\\
\indent\indent\indent\indent\indent\indent\indent \textsc{case:} $p^Ty < 1$\\
\indent\indent\indent\indent\indent\indent\indent\indent \textsc{return} $-p$\\
\indent\indent\indent\indent\indent\indent\indent \textsc{else:} $p^Ty \geq 1$\\
\indent\indent\indent\indent\indent\indent\indent\indent \textsc{call} $\OSSEPKstar(y)$\\
\indent\indent\indent\indent\indent\indent\indent\indent \textsc{case:} $\OSSEPKstar(y)$ returns separating vector $q$\\
\indent\indent\indent\indent\indent\indent\indent\indent\indent \textsc{return} $q$\\
\indent\indent\indent\indent\indent\indent\indent\indent \textsc{else:} $\OSSEPKstar(y)$ asserts $y\in K^\star$\\
\indent\indent\indent\indent\indent\indent\indent\indent\indent \textsc{return} ``$y\in Q$''\\

\noindent It remains to show that the optimization routine
$\OSOPTK$ for $K$ gives a separation routine $\OSSEPKstar$ for
$K^\star$. Suppose $y$ is given to $\OSOPTK$, which returns $k\in
K$ such that $y^Tx \leq y^Tk =:b$ for all $x\in K$. If $b\leq 1$,
then $\OSSEPKstar$ may assert $y\in K^\star$. Otherwise,
$\OSSEPKstar$ may return $k$, because $\pi_{k,1}$ (and hence
$\pi_{k,b}$) separates $y$ from $K^\star$: since $k^Ty=b > 1$, it
suffices to note that $k^Tc=c^Tk\leq 1$ for all $c\in K^\star$ by
the definition of $K^\star$ and the fact that $k\in K$.

\begin{figure}[ht]
\centering \resizebox{150mm}{!}{\includegraphics{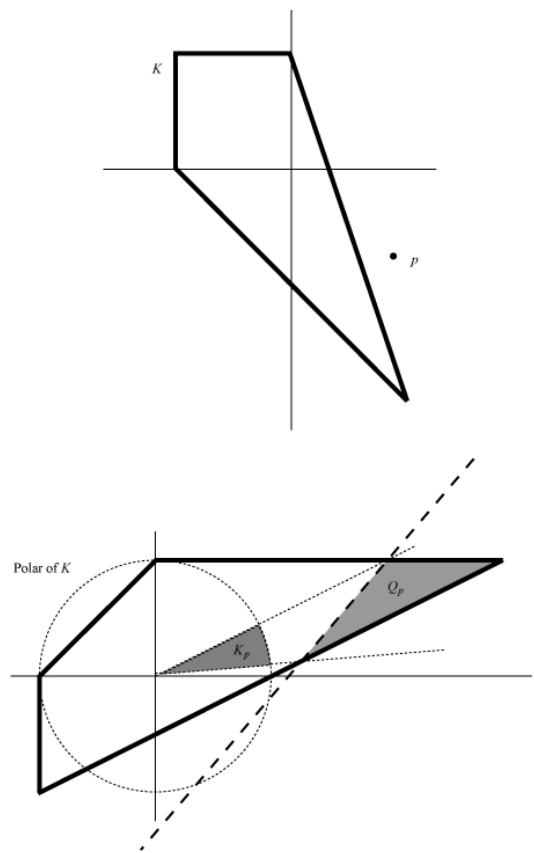}}
\caption[Relationship between convex separation methods]{The upper
picture is a set $K$ in $\mathbb{R}^2$, where  $K:=\conv\{(0,1),
(-1,1), (-1,0), (1,-2)\}$.  A point $p=(-7/8,-3/4)$ is shown.  The
polar $K^\star$ of $K$ is shown in heavy outline in the lower
picture; $K^\star = \conv \{(0,1),(-1,0),(-1,-1),(3,1)\}$.  The
set $Q_p$ is the shaded polytope, bounded by the long-dashed plane
$\{c: p^Tc =1\}$. The set $K_p$ is the shaded pie-slice and is the
radial projection of $Q_p$ onto $B_n$ (whose boundary is shown as
a short-dashed circle).  The particular $K$ and $K^\star$ are
taken from \cite{NW88}.} \label{Connection}
\end{figure}

Figure \ref{Connection} shows the relationship between the method
of Section \ref{sec_SearchHeuristic} and the above method, by
illustrating that the set $K_p$ (defined in (\ref{def_K_p})) is
just the radial projection of $Q_p$ onto $B_n$.  Using ``$A
\longrightarrow B$'' to mean ``$B$ Turing-reduces to $A$'', the
standard reduction chain is
\begin{eqnarray}
\textrm{WOPT}(K) \longrightarrow \textrm{WSEP}(K^\star)
\overset{R}{\longrightarrow} \textrm{WSEP}(Q_p) \longrightarrow
\textrm{WFEAS}(Q_p),
\end{eqnarray}
where we indicate that the middle reduction requires knowledge of
outer radius $R$.  Our reduction, which is substantially less
general, may be written
\begin{eqnarray}\label{OurTuringReductionChain}
\textrm{WOPT}(K) \longrightarrow
\textrm{WSEP}(K_p)\Bigl.\Bigr|_{(\ref{HypothesisOfDonnyLemma})}
\overset{R,
c_0}{\underset{(\ref{eqn_CenterIsConicCombinationOfNormals})}{\longrightarrow}}
\textrm{WFEAS}(K_p),
\end{eqnarray}
where the weak separation problem has been restricted (queries
must satisfy the condition (\ref{HypothesisOfDonnyLemma}) of Lemma
\ref{lem_Donny}) and we note that the rightmost link requires, in
addition to knowledge of $R$ and $c_0$, a WFEAS algorithm whose
centers satisfy (\ref{eqn_CenterIsConicCombinationOfNormals}) (see
Section \ref{sec_AnalyticCenters}).

The two reductions have similar structure; the novelty of our
reduction lies in the \emph{way} the cutting planes are generated,
which is based on the heuristic:  if the center $\omega$ satisfies
(\ref{eqn_CenterIsConicCombinationOfNormals}), then cutting near
$\omega$ pushes the new center in the direction of the normal of
the new cutting plane; but this direction is essentially $(p-k_c)$
(except with the projection onto $c$ removed).  Note that the
normal of every new cutting plane depends on the given point $p$,
whereas in the standard reduction it does not.  These attributes
tempt us to conjecture that our reduction performs ``better'' in
practice when $p\notin K$ (though not in any significant
asymptotic sense).

Note that, even though $\OSSEPKstar$, built on $\OSOPTK$, gives
deep cuts $\pi_{k,1}$, it is not known how to utilize the deep
cuts to get a polynomial-time algorithm using analytic or
volumetric centers. Our cut-generation method in Section
\ref{sec_SearchHeuristic} is capable only of giving central cuts;
but this does not, \emph{a priori}, put it at any disadvantage
(relative to the standard cut-generation method) with regard to
polynomial-time analytic or volumetric center algorithms.  In the
next section, we briefly outline how our heuristic indeed yields a
polynomial-time algorithm (justifying the Turing reduction in
(\ref{OurTuringReductionChain})).

\section{Analytic Centers}\label{sec_AnalyticCenters}

Let $P$ be the current outer approximation $P:=B_n\bigcap
\cap_{i=1}^{h}\lbrace x:a_i^T x\geq b_i \rbrace$, as described in
the second-last paragraph of Section \ref{sec_SearchHeuristic}.
Recall that we need a definition of ``center $\omega$ of $P$''
that satisfies (\ref{eqn_CenterIsConicCombinationOfNormals}).
Define the \emph{analytic center} $\omega$ of $P$ as the unique
minimizer of the real convex function
\begin{eqnarray}\label{def_F}
F(x):= -\sum_{i=1}^{h}\log(a_i^T x-b_i) - \log(1-x^T x).
\end{eqnarray}
The relation $\nabla F(\omega)=0$ gives
\begin{eqnarray}\label{eqn_defw}
\omega=\frac{1-\omega^T\omega}{2}\sum_{i=1}^{h}\frac{a_i}{a_i^T
\omega - b_i},
\end{eqnarray}
which shows that $\omega$, defined as the analytic center of $P$,
indeed satisfies (\ref{eqn_CenterIsConicCombinationOfNormals}).

Our algorithm is a modification of the one in \cite{AV95}: we use
a hypersphere for $P_0$ instead of a hyperbox (i.e. we adapt the
analysis to handle a convex quadratic constraint). We refer to
\cite{qphIoa05} for details (which rely on results in
\cite{NN94,Ren01}), but remind the reader of some of the
algorithm's characteristics. The algorithm stops when the current
outer approximation becomes either too small (volume-wise) or too
thin to contain $K_p$. For this, a lower bound $r>0$ on the radius
of the largest ball contained in $K_p$ is needed. By exploiting
the accuracy parameter $\delta$ of the weak separation problem,
such an $r$ exists (and is derived in \cite{qphIoa05}). The actual
algorithm is not as straightforward. For instance, it is a
shallow-cut algorithm ($b_i<0$) so as to keep the analytic center
of the old $P$ in the new $P$ and the new center close to the old
center. As well, cutting planes are occasionally discarded so that
$h$ does not exceed some prespecified number.

Incidentally, it is an open problem whether there exists a
polynomial-time, analytic center algorithm for the convex
feasibility problem that does not require shallow cutting. It has
been suggested by Mitchell \cite{Mit03,Mit05} that central cutting
can be used in the Atkinson-Vaidya algorithm, if certain
techniques from \cite{MT92} are employed to compute the new
analytic center. However, from correspondence with Mitchell and Ye
\cite{Ye05}, it is unclear whether modifying the Atkinson-Vaidya
algorithm in this way retains the polynomial-time convergence:
while it is clear that a new analytic center can be efficiently
computed when central cuts are used, it is not clear that all the
other delicate machinery in the convergence argument emerges
unscathed.  If central cuts may indeed be used, then the
worst-case precision requirements of our algorithm are
significantly reduced \cite{qphIoa05}.

\section{Acknowledgements}

We gratefully acknowledge an anonymous referee who pointed out the
relationship between our reduction and the standard reduction,
allowing us to streamline this paper and correctly identify its
novel contribution.  We also thank Tom Stace and Coralia Cartis
for assistance.  LMI acknowledges the support of GCHQ and ORS
(UK), and NSERC (Canada); BCT acknowledges CMI (UK); DC
acknowledges NSERC and the University of Waterloo.

\bibliographystyle{unsrt}


\end{document}